\documentclass[cits]{PoS}
\usepackage{multirow}
\title{The meson spectrum in large-$N$ QCD}

\ShortTitle{The meson spectrum in large-$N$ QCD}

\author{\speaker{Gunnar S.\ Bali}\thanks{Adjunct Faculty:
Tata Institute of Fundamental Research, Homi Bhabha Road, Mumbai 400005, India.}, Luca Castagnini, Sara Collins\\
        Institute f\"ur Theoretische Physik,  University of Regensburg,
D-93040 Regensburg, Germany\\
        E-mail: \email{gunnar.bali@ur.de}, \email{luca.castagnini@ur.de},
\email{sara.collins@ur.de}}
\author{Francis Bursa\\
SUPA, School of Physics and Astronomy, University of Glasgow,
Glasgow G12 8QQ, UK\\
E-mail: \email{francis.bursa@gla.ac.uk}}
\author{Luigi Del Debbio\\
SUPA and The Tait Institute, School of Physics and Astronomy, University of Edinburgh, Edinburgh EH9 3JZ, UK\\
E-mail: \email{luigi.del.debbio@ed.ac.uk}}
\author{Biagio Lucini\\
College of Science, Swansea University, Singleton Park, Swansea SA2 8PP, UK\\
E-mail: \email{b.lucini@swansea.ac.uk}}
\author{Marco Panero\\
Department of Physics and Helsinki Institute of Physics,
University of Helsinki, FIN-00560 Helsinki, Finland\\
E-mail: \email{marco.panero@helsinki.fi}}

\abstract{We present lattice results on the meson spectrum and decay
constants in large-$N$ QCD. The results are obtained in the quenched
approximation for $N=2,3,4,5,6,7$ and $17$ and extrapolated to $N=\infty$.
}
\FullConference{Xth Quark Confinement and the Hadron Spectrum,\\
		October 8-12, 2012\\
		TUM Campus Garching, Munich, Germany}

\begin{document}
\section{Introduction}
Quantum Chromodynamics (QCD), the theory of strong interactions
is a non-Abelian quantum field theory characterized by local $\mathrm{SU}(N)$
gauge invariance where $N=3$ denotes the number of ``colours''.
The adjoint gauge bosons (gluons) couple $n_f$ ``flavours'' of
fermionic matter fields in the fundamental representation
(quarks).

The standard non-perturbative definition of QCD is based on
lattice regularization~\cite{Wilson:1974sk}, which makes the
theory mathematically well-defined and amenable to analytical
as well as numerical studies. Theoretical progress, algorithmic
innovation and ever more powerful computers have allowed some
teams to derive non-perturbative low energy properties like the spectrum
of QCD with $n_f=2$, $n_f=2+1$ and $n_f=2+1+1$ sea quarks
at realistic values of the physical parameters, see, e.g.,
ref.~\cite{Fodor:2012gf} for a recent review.

A different non-perturbative approach to QCD is based on an expansion
in powers of $1/N$ of the inverse number of colour charges
at fixed $n_f$~\cite{largeN}.
When $N$ is taken to infinity, and the gauge coupling $g$ is sent to zero,
keeping the product $g^2 N$ as well as $n_f$ fixed
(the 't Hooft limit), the theory reveals striking
mathematical simplifications, see refs.~\cite{reviews} for recent reviews.
For instance, all amplitudes of physical processes are determined
by a particular subset of Feynman diagrams (planar diagrams),
the low-energy spectrum consists of stable meson and glueball states
and the scattering matrix becomes trivial.
One may study the physical $N=3$ case, expanding around the large-$N$
limit in terms of $1/N$. Interestingly,
the non-flavour-singlet spectra of QCD with sea quarks and quenched QCD
agree within 10~\%~\cite{Fodor:2012gf}, which may indicate both $n_f/N$ and
$1/N^2$ corrections to be small.

The large-$N$ limit also plays a vital role in the chiral effective
theory approach where the $N$-dependence of
low-energy constants is known~\cite{Sharpe:1992ft} and,
within this framework, in studies of properties of unstable resonances,
see, e.g., refs.~\cite{Pelaez:2006nj,Nieves:2009ez}.
(Un)fortunately, even in the large-$N$ limit
QCD is far from trivial.

Another non-perturbative approach --- that, unlike lattice regularization,
does not break the Euclidean spacetime symmetry --- to low-energy
properties of strongly coupled non-Abelian gauge theories is based on the
conjectured correspondence between (supersymmetric) large-$N$
gauge and string theories in the classical gravity limit in
an anti-de-Sitter spacetime
(AdS/CFT correspondence)~\cite{gaugestringduality}. During the
last decade, many studies have used techniques based on this
correspondence to construct models which reproduce the main features
of the meson spectrum of QCD~\cite{Erdmenger:2007cm}.

Recently, the dependence of various quantities on
$N$ was studied in lattice simulations. For instance, 
pseudoscalar and vector meson masses (among other observables)
were determined in
refs.~\cite{DelDebbio:2007wk, Bali:2008an, DeGrand:2012hd, Hietanen:2009tu}.
Here we improve upon and extend these studies, reducing the
quark masses and increasing $N$, the statistics, the number
of states studied and the volume.

In view of the above discussion, it is important to determine
the meson spectrum of large-$N$ QCD to constrain effective
field theory parameters and to enable comparison with AdS/CFT and
AdS/QCD predictions.
We perform our simulations in the quenched approximation to
QCD, neglecting sea quark loops.
Therefore, we only encounter $1/N^2$ corrections to the large-$N$
limit, rather than $n_f/N$ corrections. This allows
for a more constrained $N\rightarrow\infty$ extrapolation,
at the same time reducing the computational effort.
We remark, however, that the naive cost of including sea quarks
into the update only scales like $N^2$ while the pure gauge
operations scale like $N^3$: in the large-$N$ limit not
only the quenched theory becomes unitary and identical
to the un-quenched theory but so also does the computational effort,
which is quite substantial at $N=17$.

Finally, we aim at clarifying a discrepancy between the results of
refs.~\cite{DelDebbio:2007wk, Bali:2008an,DeGrand:2012hd}, which at
large $N$ favour a value of the vector meson mass close to that of
real-world QCD, and those obtained in ref.~\cite{Hietanen:2009tu},
reporting a value approximately twice as large. 

\section{Simulation details}
Our simulation strategy is to
tune the lattice couplings,
keeping the square root of the
string tension $\sqrt{\sigma}a\approx 0.2093$ 
in lattice units $a$ fixed.
We employ the Wilson gauge and fermionic action and
are not yet in the position to perform a
continuum limit extrapolation. 
Lattice artefact terms will have the same functional
large-$N$ scaling as the dominant continuum limit terms and hence
basically should not affect the size of $1/N^2$ corrections.
Our experience
with the present action~\cite{DelDebbio:2007wk,Bali:2008an}
leads us to expect systematic errors on mass ratios
of about 5~\% at the present lattice spacing.

We define the decay constant of a meson $X$ in the large
$N$ limit as
\begin{equation}
F^{\infty}_X=\lim_{N\rightarrow\infty}
\sqrt{\frac{3}{N}}F_X\,,
\end{equation}
where $F_X=f_X/\sqrt{2}$. We distinguish between
$F_{\pi}\approx 92$~MeV at physical quark masses and
$F=F_{\pi}(m_q=0)$. 

Using the \emph{ad hoc} value $\sigma=1$~GeV/fm, 
our lattice spacing corresponds to $a\approx 0.093$~fm
or $a^{-1}\approx 2.1$~GeV.
Strictly speaking, we can only predict ratios of dimensionful quantities.
In the real world where experiments are performed,
$n_f>0$, $N=3\neq\infty$ and even the string tension is
not well defined. This means that any absolute scale
setting in physical units
will be arbitrary and is just meant as a rough guide.
Nevertheless, we notice that
other ways of setting the scale give similar results, indicating
that the $N=\infty$ world is not far removed from $N=3$ QCD with
sea quarks. For instance,
in the chiral limit
we find $F^{\infty}= 0.22(2)\sqrt{\sigma}=96(9)$~MeV,
in qualitative agreement with the
real QCD value~\cite{Colangelo:2010et} $F=85.9(1.2)$~MeV.
Moreover, we obtain $m_{\rho}=1.638(7)\sqrt{\sigma}=728(3)$~MeV
at large $N$,
quite close to the experimental $\rho$-meson mass
of 775~MeV. This is remarkable, in particular since this resonance has a
decay width of almost 150~MeV.

The string tension was computed
in ref.~\cite{Lucini:2005vg} for $N=2, 3, 4, 6$ and $8$
and in ref.~\cite{Lucini:2012wq} for $N=5$ and $7$.
For $N=17$ no value is known and our lattice volume (see below)
is too small for a reliable determination from torelon correlators.
Therefore, we estimate
\begin{equation}
\label{lambdalattice}
\Lambda \approx a^{-1} \exp\left[-\frac{1}{2 \beta_0
\alpha(a^{-1})}\right]\left[ \beta_0 \alpha(a^{-1}) \right]^{-\frac{\beta_1}{2 \beta^2_0}} \left[1 + \frac{1}{2 \beta_0^3}\left( \beta_1 ^2 - \beta_2^L \beta_0 \right) \alpha(a^{-1}) \right] 
\end{equation}
and extrapolate the $\Lambda$-parameter obtained
at $N\leq 8$ for $a\sqrt{\sigma}=0.2093$
as a polynomial in $1/N^2$ to $N=17$. The lattice
coupling is defined as $\beta=2N^2/\lambda=N/(2\pi\alpha)$,
where $\lambda=Ng^2$ is the 't~Hooft parameter in the lattice scheme.
We obtain the central value
$\beta_{17}=208.45^{+59}_{-29}$ from a $1/N^2$ fit to the $N\geq 6$ data, with
systematics estimated by varying the fit range and allowing for a quartic
term. Our $\beta$-values and simulated volumes are summarized in
table~\ref{parameters}. Note that the lattice 't Hooft couplings deviate
by $1/N^2$ terms from a constant along our trajectory of fixed
$\sigma a^2$. We remark that incidentally
ref.~\cite{Hietanen:2009tu}
simulated $\mathrm{SU}(17)$ at $\beta=208.08$ which is
almost identical to our value $\beta_{17}=208.45$, ruling out
the hypothesis that our $N<17$ simulation points are not
in the ``continuum phase of the large-$N$ theory''~\cite{Hietanen:2009tu}.
In fact at our lattice spacing we will find the $7\geq N\geq 3$ spectra
to be in almost perfect agreement with the $N=17$ results.

Different lattice sizes are investigated to exclude finite volume
effects spoiling the large-$N$ extrapolation,
in particular at the lighter quark masses. As expected, these
become irrelevant at large $N$, thereby justifying the use of a
relatively small lattice at $N=17$.

The so-called hopping parameter $\kappa$ is related to the
lattice quark mass $m_q$ via
\begin{equation}
\label{eq:bare_mass}
am_q=\frac{1}{2}\left(\frac{1}{\kappa}-\frac{1}{\kappa_c}\right)\,.
\end{equation}
$\kappa_c$ denotes the critical value, corresponding to a massless quark.
The additive constant is given by $\kappa_c^{-1}=8+{\mathcal O}(\lambda)$
and we will determine this non-perturbatively.
The $\kappa$-values shown in table~\ref{parameters} were selected
to keep one set of pion masses approximately constant across the
different $\mathrm{SU}(N)$ theories. We vary the ``pion'' mass
down to $m_\pi/\sqrt{\sigma}  \approx 0.5$ for groups with $N \geq 5$,
and to $m_\pi/\sqrt{\sigma}  \approx 0.75$ for $N<5$. We also simulated
a smaller quark mass for $\mathrm{SU}(N<5)$ but found significant numbers
of ``exceptional configurations"~\cite{Bardeen:1997gv} (up to 15~\% of
the total); we leave these data out of this work.
For $N = 5$, at the lowest quark mass, only two
exceptional configurations were encountered that we removed from the analysis.

Our code is based on the Chroma suite~\cite{Edwards:2004sx},
which we have adapted to work for generic $N$ values.
We compute correlation matrices between differently
smeared interpolators, allowing us not only to extract the
ground states but also giving us access to excitations in
many channels. Details of the analysis can be found in
ref.~\cite{refnew}.

{\small\TABULAR[t]{|c|c|c|c|c|c|}
{\hline 
$N$  & $N_s^3 \times N_t$ & $\beta$ & $\lambda$&$10^5\kappa$ & $n_{\mathrm{conf}}$ \\
\hline
$2$ & $16^3 \times 32$ & \multirow{3}{*}{2.4645}  &\multirow{3}{*}{3.246}& 14581,  14827,  15008,  15096 &  400 \\ 
 & $24^3 \times 48$ &  & &14581,  14827,  15008,  15096,  15195.9 ,15249.6   &  200 \\ 
  & $32^3 \times 64$ &  & & 14581,  14827,  15008,  15096,  15195.9 ,15249.6   &  100 \\ 
  \hline
  $3$ & $16^3 \times 32$ &\multirow{3}{*}{6.0175} &\multirow{3}{*}{2.991}&  15002,  15220, 15380, 15458   &  200\\ 
 & $24^3 \times 48$ &  &&  15002,  15220, 15380, 15458, 15563.8, 15613   & 200 \\ 
  & $32^3 \times 64$ &  &&   15002,  15220, 15380, 15458, 15563.8, 15613  &  100 \\ 
\hline
  $4$ & $16^3 \times 32$ &\multirow{2}{*}{11.028} &\multirow{2}{*}{2.902}& 15184, 15400, 15559, 15635   &  200\\ 
 & $24^3 \times 48$ &  &&   15184, 15400, 15559, 15635, 15717.3, 15764    &  200\\ 
 \hline
  $5$ & $16^3 \times 32$ &\multirow{2}{*}{17.535}&\multirow{2}{*}{2.851}& 15205, 15426, 15592, 15658   & 200 \\ 
 & $24^3 \times 48$ &  & &   15205, 15426, 15592, 15658, 15754.8, 15835.5  &  200\\ 
\hline
  $6$ & $16^3 \times 32$ &\multirow{2}{*}{25.452} &\multirow{2}{*}{2.829}&15264, 15479, 15636, 15712     & 200 \\ 
 & $24^3 \times 48$ &  &&  15264, 15479, 15636, 15712, 15805.1, 15884.5     &  200 \\ 
\hline
  $7$ & $16^3 \times 32$ &\multirow{2}{*}{34.8343}&\multirow{2}{*}{2.813}&  15281.6, 15496.7, 15654.7, 15733.9  &  200\\ 
 & $24^3 \times 48$ &  & & 15281.6, 15496.7, 15654.7, 15733.9, 15827.3, 15906.2    &  200 \\ 
\hline
  $17$ & $12^3 \times 24$ & $208.45$ &2.773&  15298, 15521, 15684, 15755,  15853.1,  15931 &  80\\ 
\hline
}
{{\normalsize Simulation parameters: $n_{\mathrm{conf}}$
is the number of configurations analysed
for each set of parameters. All configurations were separated by 200
combined heatbath and overrelaxation Monte Carlo sweeps and found to
be effectively statistically independent.
\label{parameters}}}}
\section{Results}
We employ the lattice quark mass $m_{\mathrm{PCAC}}$, defined through the
axial Ward identity, as our reference mass,
avoiding the additive renormalization ($\kappa_c^{-1}$)
of the quark mass $m_q$ defined in eq.~(\ref{eq:bare_mass}).
These two masses are related
by a combination of renormalization constants,
\begin{equation}
a\,m_{\mathrm{PCAC}}= \frac{Z_P}{Z_A Z_S} \left( 1 + b \,am_{\mathrm{PCAC}} + \cdots \right) \frac{1}{2} \left(\frac{1}{\kappa}-\frac{1}{\kappa_c}\right)\,,
\end{equation}
where the $b\,am$-term parameterizes the leading lattice correction.
All three fit parameters
$\kappa_c$, $b$ and ${Z_P}/({Z_A Z_S})$
are well described by constants
plus $1/N^2$-corrections.
We find the latter ratio of renormalization constants to
vary between
0.68 ($N=2$) and 0.83 ($N=17$), with the $\mathrm{SU}(3)$-value 0.75,
which is consistent with the non-perturbative result
0.81(7)~\cite{Gimenez:1998ue} obtained at $\beta=6.0$ --- close
to our value $\beta=6.0175$. Motivated by the weak $N$-dependence
of this result --- which is also supported by perturbation theory ---
we will use the non-perturbative $\mathrm{SU}(3)$-values
for the renormalization factors of quark bilinears for all gauge groups,
allowing for a $8\,\%$ systematic uncertainty, due to this approximation.
These factors are needed to determine the decay constants below.
\FIGURE[t]
{
\includegraphics[width=.47\textwidth,clip]{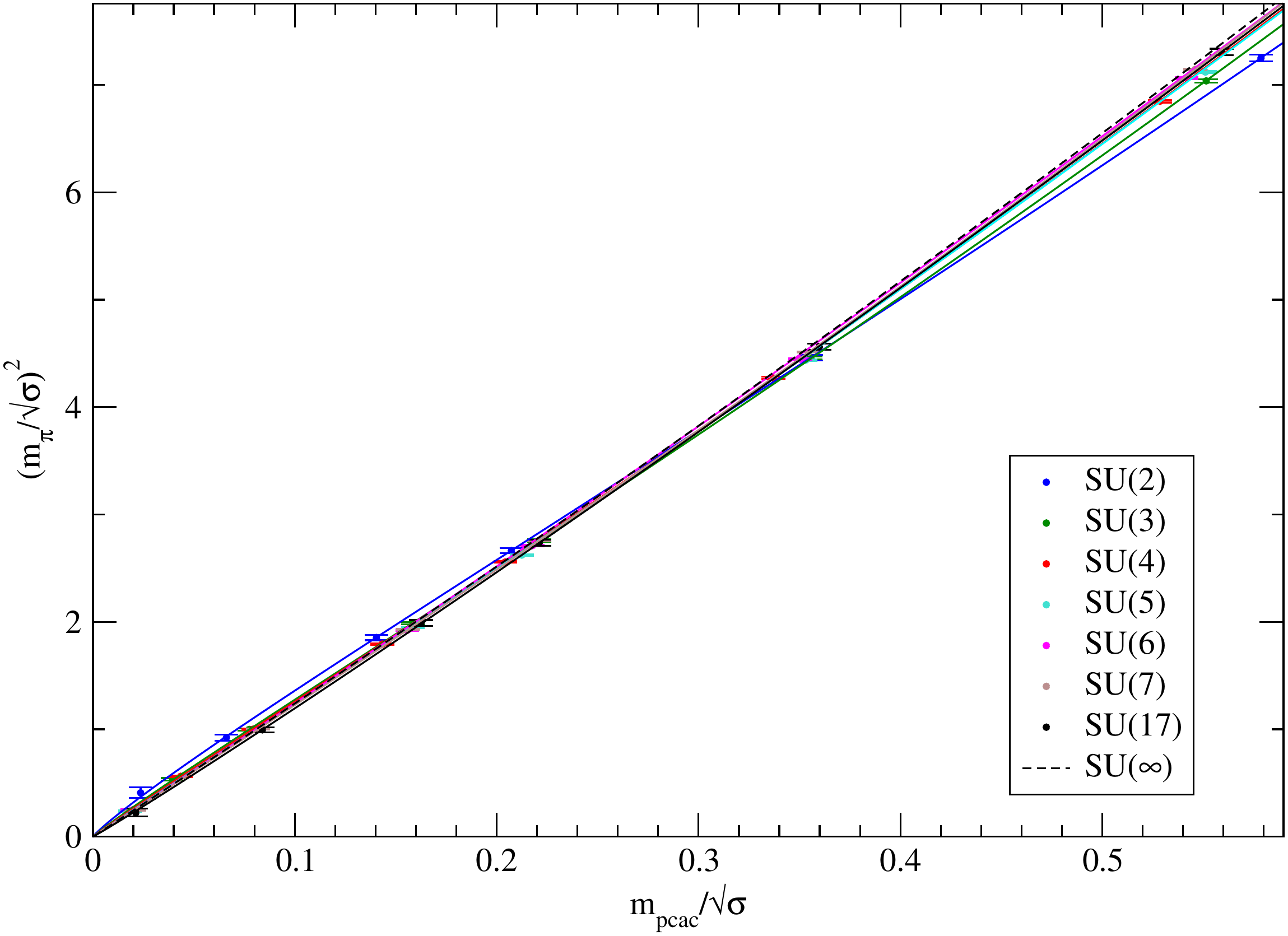}~~
\includegraphics[width=.49\textwidth,clip]{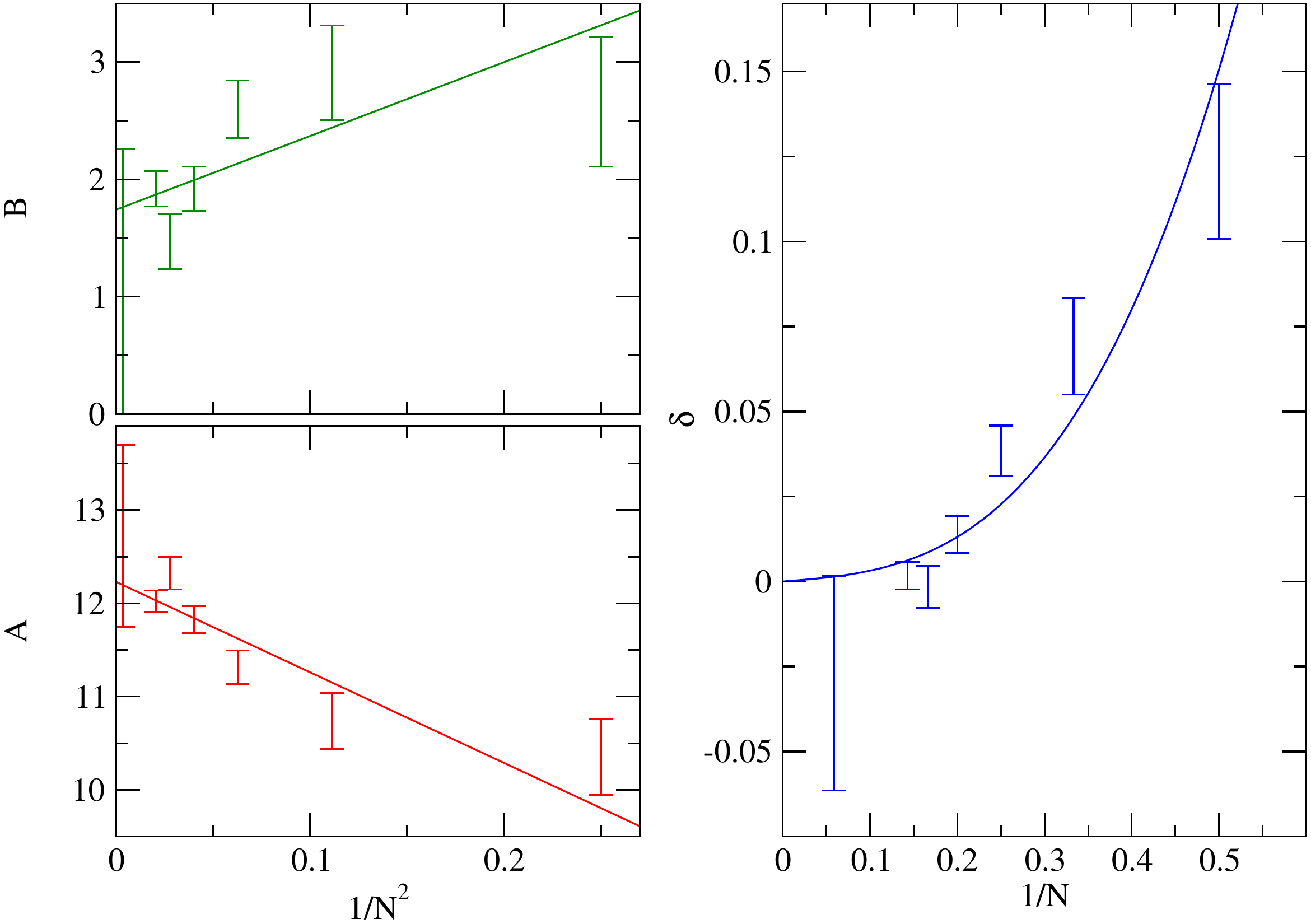}
\caption{Pion mass vs.\ PCAC mass
eq.~(\protect\ref{piondelta2}) (left). $N$ dependence of the
fit parameters~eq.~(\protect\ref{eq:fit}) (right).\label{fig:pion}}}

Next, we determine the dependence of the pseudoscalar mass
$m_{\pi}$ on the quark mass $m_{\mathrm{PCAC}}$:
\begin{equation}\label{piondelta2}
\frac{m_\pi^2}{\sigma} = A \left( \frac{m_{\mathrm{PCAC}}}{\sqrt{\sigma}} \right)^{\frac{1}{1+\delta } }  + B\,  \frac{m_{\mathrm{PCAC}}^2}{\sigma}\,.
\end{equation}
A quenched chiral log is expected at small $N$-values,
parameterized by $\delta$.
We also include a subleading term to prevent interference
between the larger mass data and the chiral log. The fits
for the different $N$ are depicted in figure~\ref{fig:pion}.
$\delta$ is expected to
be suppressed by a factor~\cite{Sharpe:1992ft} $1/N$,
with $1/N^3$ corrections.
We find the parameter values
\begin{equation}
\label{eq:fit}
A= 12.23(0.10)  - \frac{9.7(1.6)}{N^2}\,,\quad
B= 1.74 (0.13) +  \frac{6.3(2.2)}{N^2}\,,\quad
\delta = \frac{0.021(19)}{N} + \frac{1.12(21)}{N^3}\,,
\end{equation}
see the right panel of figure~\ref{fig:pion}.
$\delta\gtrsim 0.05$ at $N=3$ is coherent
with expectations but the $1/N$ coefficient is
statistically compatible with zero. In principle,
the determination of $\delta$ may be obscured by the possibility
of a non-zero pion mass at $m_{\mathrm{PCAC}}=0$,
at finite lattice spacings. However,
the linear disappearance of $m_{\pi}^2\propto m_{\mathrm{PCAC}}$
for $N\geq 5$ indicates chiral symmetry
to be broken only mildly.

We choose to parameterize the quark mass dependence of
all our results through $m_{\mathrm{PCAC}}$ which can be
determined more precisely and reliably than the pseudoscalar mass
$m_{\pi}$. The relation $m_{\pi}^2(m_{\mathrm{PCAC}})$ above enables
the translation of a functional dependence on $m_{\mathrm{PCAC}}$ into
a dependence on $m_{\pi}$. Nevertheless, we quote
the $N\rightarrow\infty$ result
\begin{equation}
\frac{m_{ \rho} (m_\pi) }{m_\rho(0) } =  1+ 0.375(64)\left( \frac{m_\pi}{ m_\rho(0) }\right)^2  +\cdots \,,
\end{equation}
to allow for a direct comparison with the prediction, e.g.,
of ref.~\cite{Babington:2003vm}.

\FIGURE[t]
{
\includegraphics[width=.76\textwidth,clip]{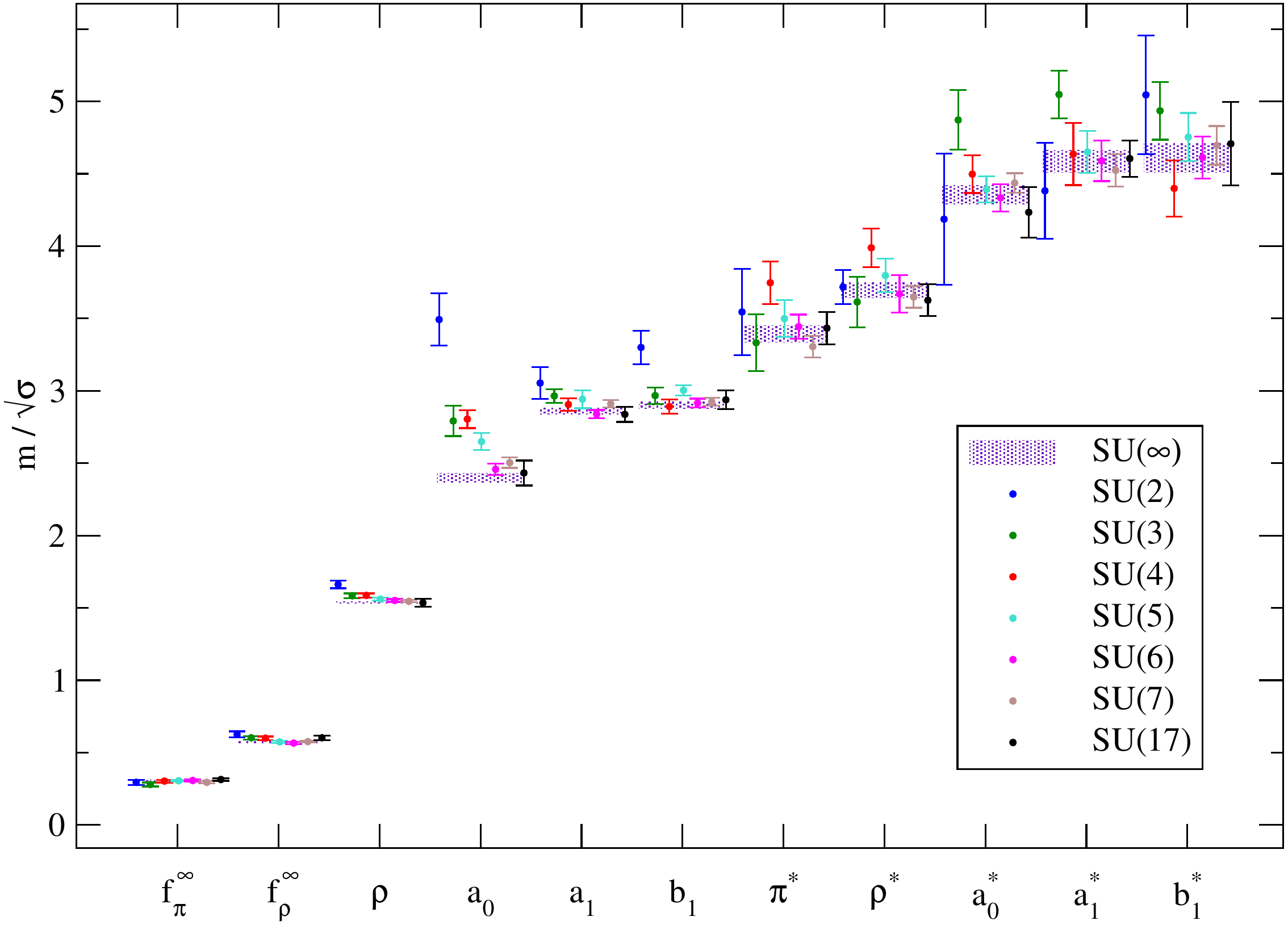}
\caption{The spectrum, extrapolated to the chiral limit.
The error bands correspond to $N=\infty$.
\label{fig:global}}}

We display the chirally extrapolated spectrum
in figure~\ref{fig:global} for the different
$\mathrm{SU}(N)$ groups. The $N\rightarrow\infty$
values are shown as horizontal error bands.
Note that the $N=17$ values are perfectly consistent
with the $N<17$ results, ruling out the twice as large
$\rho$-meson mass obtained at almost the same
lattice coupling in ref.~\cite{Hietanen:2009tu}.
This may be an artefact of the method used in that reference,
e.g., due to excited state pollutions.

The figure does not only give the masses of the
excited and ground state mesons but
also illustrates the decay constants
$f_{\pi}$ and $f_{\rho}$ 
to show the expected scaling behaviour
$\propto\sqrt{N}$, with small corrections.
The analysis of the scalar ($a_0$) 
correlation function at low $N$
is complicated by the presence of ghost
states, due to the unitarity violation of the
quenched model. Subtracting these contributions
results in large errors. 

Of phenomenological interest is not only the spectrum at
$m_q=0$ but are also the spectra at $m_q=m_{ud}$ and
at $m_q=m_s$ where $m_{ud}$ and $m_s$ denote the
physical (isospin-averaged) light quark and strange quark
masses, respectively. We fix the former, imposing
the value~\cite{Colangelo:2010et}
\begin{equation}
\frac{F_{\pi}(m_{ud})}{F}=1.073(15)\,,
\end{equation}
at $N=3$, keeping $m_{ud}/\sqrt{\sigma}$ constant
for $N\neq 3$.
The renormalization constant and $N$-dependence cancel
from the ratio.
The strange quark mass is obtained by fixing the ratio
of a (fictitious) strange-antistrange pion over
the $\phi(1020)$ vector particle at $N=3$ to the experimental value
\begin{equation}
\frac{m_{\pi}(m_s)}{m_{\rho}(m_s)}=\frac{686.9}{1019.5}\,,
\end{equation}
where $(m^2_{K^{\pm}}+m^2_{K^0}-m^2_{\pi^{\pm}})^{1/2}\approx 686.9\,\mathrm{MeV}$.
We display the results at the different quark masses
in table~\ref{tab:largeNsummary}. The normalization
in units of $F^{\infty}$ should be particularly useful for chiral
perturbation theory applications~\cite{Pelaez:2006nj,Nieves:2009ez}.
\TABULAR[t]{|c|c|c|c|c|c|c|c|} 
{\hline 
\cline{3-8} \multicolumn{2}{|c}{}  &\multicolumn{3}{|c|}{  ${m_\infty}/{ \sqrt{\sigma}}$}&\multicolumn{3}{|c|}{${m_\infty }/{F^\infty} $ }\\ \hline
Particle &$J^{PC}$&$m_q = 0$ & $m_q=m_{ud}$ &$m_q=m_{s}$ &$m_q = 0$ &$m_q=m_{ud}$&$m_q=m_{s}$\\ \hline
$\pi$ &$0^{-+}$  & 0 & 0.417(100) & 1.62(10) & 0 & 1.92(46) & 7.46(48) \\ \hline
$\rho$ &$1^{--}$ & 1.5382(65) & 1.6382(66) & 1.9130(79) & 7.08(10) & 7.54(11) & 8.80(13) \\ \hline
$a_0$ &$0^{++}$  & 2.401(31) & 2.493(31) & 2.755(32) & 11.04(21) & 11.47(22) & 12.67(23) \\ \hline
$a_1$ &$1^{++}$ & 2.860(21) & 2.938(21) & 3.158(22) & 13.16(21) & 13.51(21) & 14.53(23) \\ \hline
$b_1$ &$1^{+-}$  & 2.901(23) & 2.978(23) & 3.197(23) & 13.35(21) & 13.70(22) & 14.71(23) \\ \hline
$\pi^*$ &$0^{-+}$ & 3.392(57) & 3.462(57) & 3.659(58) & 15.61(34) & 15.93(35) & 16.83(36) \\ \hline
$\rho^*$ &$1^{--}$ & 3.696(54) & 3.756(54) & 3.928(54) & 17.00(34) & 17.28(35) & 18.07(36) \\ \hline
$a_0^*$ & $0^{++}$ & 4.356(65) & 4.420(65) & 4.603(66) & 20.04(41) & 20.33(41) & 21.18(42) \\ \hline
$a_1^*$ &$1^{++}$  & 4.587(75) & 4.646(75) & 4.816(77) & 21.10(46) & 21.38(46) & 22.15(47) \\ \hline
$b_1^*$ &$1^{+-}$ & 4.609(99) & 4.673(99) & 4.85(10) & 21.20(54) & 21.50(55) & 22.33(56) \\ \hline
$f_{\pi}^{\infty}$ & ---  & 0.3074(43) & 0.3271(44) & 0.3784(56) & $\sqrt{2}$ & 1.505(29) & 1.741(36) \\ \hline
$f_{\rho}^{\infty}$ & ---  & 0.5721(49) & 0.5855(50) & 0.6196(64) & 2.632(43) & 2.694(44) & 2.850(50) \\ \hline
}{The $N=\infty$ meson spectrum and decay constants
in units of the square root of the
string tension $\sqrt{\sigma}$ and in units of the (normalized)
chiral pion decay constant $F_{\infty}= F\sqrt{3/N}$
for three different values of the quark mass.
A systematic error of 5~\% needs to be added, due to the missing
continuum limit extrapolation. Because of the non-perturbative
$N=3$ rather than $N=\infty$ renormalization, an extra 8~\% error should be
added to the last three columns and to the last two rows of the
table, with the exception of the $f_{\pi}^{\infty}/F^{\infty}$-ratios
where this factor cancels.\label{tab:largeNsummary}}

\section{Summary}
We have determined the decay constants as well as the ground and first excited
state masses of mesons in the large-$N$ limit of QCD by lattice simulation
of the $N=2, 3, 4, 5, 6, 7$ and 17 quenched theories. In almost all
but the scalar channels $1/N^2$ corrections are found to be tiny for $N\geq 3$.
In a forthcoming publication~\cite{refnew} we will compare our
findings to model expectations.
We find the scalar to be by factors of
about 1.5 heavier than the vector particle at any quark mass
smaller than the strange quark, which is of particular relevance
to the phenomenology of scalar mesons.

\acknowledgments
\begin{sloppypar}
This work is partially supported by the EU ITN STRONGnet (grant 238353),    
by the German DFG (SFB/TRR 55), by the UK STFC (grants
ST/G000506/1 and ST/J000329/1)
and by the Academy of Finland, project 1134018.
B.\ Lucini is a Royal Society University Research Fellow.
The simulations were performed on the Regensburg
iDataCool cluster, at LRZ Munich, at the Finnish IT Center for
Science (CSC), Espoo, and on the Swansea BlueGene/P system
(part of the DiRAC Facility, jointly funded by STFC, the Large
Facilities Capital Fund of BIS and Swansea University).
We thank S.~Solbrig for assistance.\end{sloppypar}

\end{document}